\documentclass[%
 reprint,
 superscriptaddress,
 showpacs,preprintnumbers,
 nofootinbib,
 amsmath,amssymb,
 aps,
 prd,
]{revtex4-1}

\usepackage{dcolumn}
\usepackage{bm}
\usepackage{xcolor}

\newcommand{\beq}{\begin{equation}}
\newcommand{\eeq}{\end{equation}}
\newcommand{\diff}{\textrm{d}}

\begin{document}

\title{Electromagnetic self-force in curved spacetime:\\ New insights from the Janis-Newman algorithm}

\author{Matteo Broccoli}
 \email{matteo.broccoli2@studio.unibo.it}
 \affiliation{%
 Max-Planck-Institut f\"ur Gravitationsphysik (Albert-Einstein-Institut),\\
 Am M\"uhlenberg 1, 14476 Potsdam, Germany
}%
\affiliation{%
 Universit\`a degli Studi di Bologna, Dipartimento di Fisica e Astronomia,\\
 Viale Berti Pichat 6/2, 40127 Bologna, Italy
}%

\author{Adriano Vigan\`o}
 \email{adriano.vigano@aei.mpg.de}
\affiliation{%
 Max-Planck-Institut f\"ur Gravitationsphysik (Albert-Einstein-Institut),\\
 Am M\"uhlenberg 1, 14476 Potsdam, Germany
}%

\begin{abstract}

We present an original approach to compute the electromagnetic self-force acting on a static charge in Kerr spacetime.
Our approach is based on an improved version of the Janis-Newman algorithm and extends its range of applicability.
It leads to a closed expression which generalizes the existing one and, since it does not involve the electromagnetic potential, it simplifies the calculation of the self-force.

\end{abstract}


\maketitle

\section{Introduction}

A charged or massive particle in curved spacetime generates a field that, due to the spacetime curvature, acts on the particle itself. 
The force that it produces is called the \emph{self-force}~\cite{Poisson:2011nh}.
The electromagnetic and gravitational self-forces are relevant in the gravitational two-body problem in the extreme-mass-ratio regime and in the study of gravitational waves~\cite{Barack:2018yvs}.

In the cases of a static charge in Schwarzschild and in Kerr spacetime, the electromagnetic self-force is known in closed form~\cite{Smith:1980tv,Leaute:1982sm}.
However, the computation requires an analytic expression for the vector potential and is highly involved~\cite{Linet:1976sq,Leaute:1977zz}.
In Kerr spacetime, the calculation is carried out putting the charge on the symmetry axis of the black hole, so that the self-force for a static charge in a generic spacetime position is still unknown.

Except for these particular cases, the self-force can be computed only in a perturbative regime, due to the complexity of the effects involved in the phenomenon (this is especially true in the case of the gravitational self-force).
For this reason, it is desirable to find new methods that lead to closed expressions.

We present an approach based on the Janis-Newman algorithm, that was introduced to transform static solutions of Einstein field equations into rotating ones~\cite{Newman:1965tw,Newman:1965my}.
It has also been successfully applied to the scalar~\cite{Yazadjiev:1999ce,Erbin:2015pla} and the electromagnetic vector fields~\cite{Erbin:2014aya};
there is no evidence that the algorithm should work also for a derived quantity, like the self-force.

Nevertheless, we improve the algorithm to apply it to the electromagnetic self-force acting on a static charge in Schwarzschild spacetime and we successfully recover the corresponding force in Kerr spacetime.
Thus, we not only simplify the computation of the self-force, but we also extend the known solution to an expression which is valid for a generic spacetime position of the particle outside the ergosphere.

\section{The Janis-Newman algorithm}

Here we introduce the algorithm by showing how it transforms the Schwarzschild metric into the Kerr metric.
For a more general discussion on the Janis-Newman technique we refer to the original paper~\cite{Newman:1965tw} and to the review~\cite{Erbin:2016lzq}.

Given the Schwarzschild metric
\beq
{\diff s}^2 = - h(r^\prime) {\diff t^\prime}^2 + h(r^\prime)^{-1} {\diff r^\prime}^2 + g(r^\prime) {\diff\Omega}^2 ,
\eeq
where $h(r^\prime) = 1-2M/r^\prime$, $g(r^\prime) = r^{\prime 2}$ and ${\diff\Omega}^2={\diff\theta}^2+\sin^2\theta{\diff\phi}^2$ ,
we perform a change of coordinates to the retarded Eddington-Finkelstein null coordinate
\beq
\label{advanced}
\diff u^\prime = \diff t^\prime - \frac{\diff r^\prime}{1 - 2M/r^\prime} ,
\eeq
so that
\beq
\label{ef}
{\diff s}^2 = -h(r^\prime) {\diff u^\prime}^2 - 2\diff u^\prime \diff r^\prime + g(r^\prime) {\diff\Omega}^2 .
\eeq
This transformation is necessary, since the algorithm does not work in Schwarzschild coordinates.

Now let $u^\prime,r^\prime \in \mathbb{C}$ and replace the functions $h(r^\prime)$ and $g(r^\prime)$ with new real-valued functions $\tilde{h}(r^\prime,\bar{r}^\prime)$ and $\tilde{g}(r^\prime,\bar{r}^\prime)$, which depend on the complex coordinate $r^\prime$ and on its complex conjugate $\bar{r}^\prime$,
according to the rules
\begin{subequations}
\label{rules1}
\begin{align}
\label{rule1_r-1}
\frac{1}{r^\prime} & \longrightarrow \frac{\operatorname{Re}r^\prime}{|r^\prime|^2} , \\
\label{rule1_r2}
r^{\prime 2} & \longrightarrow |r^\prime|^2 ,
\end{align}
\end{subequations}
so that
\begin{subequations}
\begin{align}
h(r^\prime) \longrightarrow \tilde{h}(r^\prime,\bar{r}^\prime) &= 1 - \frac{2M\operatorname{Re}r^\prime}{|r^\prime|^2} , \\
g(r^\prime) \longrightarrow \tilde{g}(r^\prime,\bar{r}^\prime) &= |r^\prime|^2 .
\end{align}
\end{subequations}
We perform the complex change of coordinates
\begin{subequations}
\label{change}
\begin{align}
u^\prime & = u + ia\cos\theta , \\
r^\prime & = r - ia\cos\theta ;
\end{align}
\end{subequations}
adopting the Giampieri prescription~\cite{Giampieri:1990}
\beq
i \diff \theta = \sin\theta \diff \phi ,
\eeq
the differentials transform as
\begin{subequations}
\label{presc}
\begin{align}
\diff u^\prime & = \diff u - a\sin^2\theta \diff\phi , \\
\diff r^\prime & = \diff r +a\sin^2\theta \diff\phi .
\end{align}
\end{subequations}
Substituting~\eqref{change} and~\eqref{presc} into the metric~\eqref{ef} and defining $\rho^2=r^2+a^2\cos^2\theta$ we finally find
\beq
\label{kerr}
\begin{split}
{\diff s}^2 & = -\biggl(1-\frac{2Mr}{\rho^2}\biggr) {\diff u}^2 - 2 \diff u \diff r + \rho^2 {\diff\theta}^2 \\
& \quad - \frac{4Mr}{\rho^2} a\sin^2\theta \, \diff u \diff\phi + 2a\sin^2\theta \diff r \diff\phi \\
& \quad + \biggl[ \biggl(1+\frac{2Mr}{\rho^2}\biggr)a^2\sin^2\theta + \rho^2 \biggr] \sin^2\theta {\diff\phi}^2 ,
\end{split}
\eeq
which is precisely the Kerr metric in retarded Kerr coordinates.

With the infinitesimal change of coordinates~\cite{Wiltshire:2009zza}
\begin{subequations}
\label{bl}
\begin{align}
\diff u & = \diff t - \frac{r^2+a^2}{\Delta} \diff r , \\
\diff\phi & = -\diff\tilde{\phi} - \frac{a}{\Delta} \diff r ,
\end{align}
\end{subequations}
with $\Delta=r^2-2Mr+a^2$, we translate~\eqref{kerr} into the more familiar Boyer-Lindquist form
\beq
\begin{split}
{\diff s}^2 & = -\biggl(1-\frac{2Mr}{\rho^2}\biggr) {\diff t}^2
+ \frac{\rho^2}{\Delta} {\diff r}^2 \\
& \quad - \frac{4Mr}{\rho^2} a\sin^2\theta \diff t \diff\tilde{\phi}
+ \rho^2 {\diff\theta}^2 \\
& \quad + \biggl(r^2 + a^2 + \frac{2Mr}{\rho^2} a^2 \sin^2\theta\biggr)\sin^2\theta {\diff\tilde{\phi}}^2 .
\end{split}
\eeq

\section{The self-force of a static charge in Kerr spacetime}

\subsection{Self-force in Schwarzschild spacetime}

Let us consider now a static particle with charge $e$ located outside the Schwarzschild black hole at radial coordinate $r^\prime$. The first closed form for the self-force $f^\mu$ acting on such a particle was found in~\cite{Smith:1980tv}.
In Schwarzschild coordinates~\cite{Burko:1999zy,Poisson:2011nh} only the radial component of the force is non-zero
\beq
\label{force-schwarz}
f^{r^\prime} = \frac{Me^2}{r'^3} \biggl(1-\frac{2M}{r'}\biggr)^{1/2} .
\eeq
Since the Janis-Newman procedure does not work in Schwarzschild coodinates, we convert the expression~\eqref{force-schwarz} into Eddington-Finkelstein coordinates~\eqref{advanced} obtaining two non-zero components:
\begin{subequations}
\label{up-index}
\begin{align}
f^{u'} & = -\frac{Me^2}{r'^3} \biggl(1-\frac{2M}{r'}\biggr)^{-1/2} , \\
f^{r'} & = \frac{Me^2}{r'^3} \biggl(1-\frac{2M}{r'}\biggr)^{1/2} .
\end{align}
\end{subequations}
Defining $\mathbf{f}:=f_\mu{\diff x}^\mu$ and lowering the indices of~\eqref{up-index} with the Eddington-Finkelstein metric, we have
\beq
\label{force-ef}
\mathbf{f} = \frac{Me^2}{r'^3} \biggl(1-\frac{2M}{r'}\biggr)^{-1/2}  \diff r' .
\eeq
This expression is our starting point for the application of the Janis-Newman algorithm.

\subsection{Self-force in Kerr spacetime}

The first step of the algorithm requires to complexify the Eddington-Finkelstein coordinates by keeping the force components real. However, the rules~\eqref{rules1} do not account for the powers of $r'$ which appear in~\eqref{force-ef}. Thus we adopt the following rules:
\begin{subequations}
\label{rules2}
\begin{align}
\label{rule_r-3}
\frac{1}{r^{\prime 3}} & \longrightarrow \frac{\operatorname{Re}r^\prime}{|r^\prime|^4} ,\\
\label{rule_sqrt}
\biggl(1-\frac{2M}{r^\prime}\biggr)^{-1/2} & \longrightarrow \biggl(1-\frac{2M\operatorname{Re}r^\prime}{|r^\prime|^2}\biggr)^{-1/2} .
\end{align}
\end{subequations}
The rule \eqref{rule_r-3} is motivated by the factorization $r^{\prime -3} = r^{\prime -1} r^{\prime -2}$ and the rule~\eqref{rule1_r-1} together with
\beq
\frac{1}{r^{\prime 2}} \longrightarrow \frac{1}{|r^\prime|^2} ,
\eeq
while in~\eqref{rule_sqrt} we assume that~\eqref{rule1_r-1} is still valid.

Applying the algorithm and the set of rules~\eqref{rules2} to the force~\eqref{force-ef}, we get (in Kerr coordinates)
\beq
\label{force-kerr}
\mathbf{f} = \frac{Me^2r}{\rho^4} \biggl(1-\frac{2Mr}{\rho^2}\biggr)^{-1/2} \, \bigl(\diff r + a\sin^2\theta \diff\phi\bigr) .
\eeq
We observe that a $\phi$-component appears.
This is physically meaningful, since we expect other contributions in addition to the one in the $r$-direction for a particle located in a generic position $(r,\theta,\phi)$.
Moreover, if we choose $\theta=0$ then the $\phi$ component disappears, consistently with the fact that on the symmetry axis there are no effects on the force caused by the rotation of the black hole.

We remark that in the limit $a \to 0$ of a non-rotating black hole, $\rho^2 \to r^2$ and from~\eqref{force-kerr} we correctly recover the self-force~\eqref{force-ef} in Schwarzschild spacetime.

We express the result in Boyer-Lindquist coordinates as well:
from relations~\eqref{bl} we have
\beq
\label{force-bl}
\begin{split}
\mathbf{f} & = \frac{Me^2r}{\rho^4} \biggl(1-\frac{2Mr}{\rho^2}\biggr)^{-1/2} \\
& \quad \cdot \biggl[\biggl(1-\frac{a^2\sin^2\theta}{\Delta}\biggr)\diff r + a\sin^2\theta \diff\tilde{\phi}\biggr] .
\end{split}
\eeq
It is worth calculating the absolute value of the force~\eqref{force-bl}: defining $|f|=(f_\mu f^\mu)^{1/2}$ we find
\beq
\label{module-generic}
|f| = \frac{Me^2 r}{\rho^4} = \frac{Me^2 r}{(r^2 + a^2 \cos^2\theta)^2} ,
\eeq
which, since it is invariant under coordinate transformations, is useful when comparing~\eqref{force-bl} with the result known from the literature.
In particular, setting $\theta = 0$, a straightforward calculation (more on this in the appendix) shows that~\eqref{module-generic} reduces to the absolute value of the self-force in~\cite{Leaute:1982sm}. 

\section{Conclusions}

Although the Janis-Newman algorithm has been known for more than fifty years, up to now it has been successfully applied only as a solution generating technique for the standard bosonic fields (with spin~$< 3$).
With this work, we have shown that in the case of a static charge in curved spacetime it can also be applied to a derived quantity, like the self-force which the charge is subject to. 
In particular, the algorithm leads to a result that is more general than the one already found in~\cite{Leaute:1982sm}, since it does not rely on any special choice for the position of the particle (as long as it is outside the ergosphere, but this is a necessary condition for the particle to be static).

Since we focused on the electromagnetic self-force, we cannot tell whether the success of the algorithm is due to its success with the vector potential.
Anyway we have improved the transformation rules of the Janis-Newman algorithm and extended its range of applicability.

We have greatly simplified the computation of the electromagnetic self-force in a rotating spacetime.
Our approach does not make use of the vector potential, so that one does not have to compute it;
moreover it leads by itself to a closed result.
Thus, it will be interesting to further explore this approach, for instance by applying the technique to the case of an arbitrary moving charge or to the gravitational self-force.

\acknowledgments{The authors are grateful to Hugo A.~Camargo for introducing them to the Janis-Newman algorithm and to Harold Erbin for critical comments.
They also thank Shahar Hadar and Alex J.~Feingold for stimulating and useful discussions and Fiorenzo Bastianelli, Alexander Kegeles, Axel Kleinschmidt and Stefan Theisen for critical readings of the draft of this paper.

M.B.~thanks the UniBo School of Science for the financial supporting scholarship.
He is particularly grateful to Stefan for invaluable and stimulating discussions and for his hospitality at the AEI, where this work was done.

A.V.~is supported by the European Research Council (ERC) under the European Union's Horizon 2020 research and innovation programme (``Exceptional Quantum Gravity'', grant agreement No 740209).}

\appendix*

\section{Comparison with the literature}

To check whether the algorithm is reliable we need to compare our result with the expression presented by L\'eaut\'e and Linet in~\cite{Leaute:1982sm}.
Since their result holds for a static particle on the symmetry axis, i.e.~for $r=r_0$ and $\theta=0$, and relies on a particular local coordinate system, it is useful to briefly review their setup.

They start with the Kerr metric in Boyer-Lindquist coordinates $(t,r,\theta,\tilde{\phi})$ and then introduce the Cartesian-like coordinates $(x^0, x^i)$ by:
\begin{subequations}
\label{cartesian-like-coord}
\begin{align}
x^0 & = t , \\
x^1 & = r\sin\theta\cos\tilde{\phi} , \\
x^2 & = r\sin\theta\sin\tilde{\phi} , \\
x^3 & = r\cos\theta .
\end{align}
\end{subequations}
Subsequently they define a local set of coordinates $(y^0, y^i)$ satisfying the relations
\begin{subequations}
\label{local-coord}
\begin{align}
x^0 & = \Delta_0^{-1/2} \rho_0 \, y^0 , \\
x^1 & = r_0 \rho_0^{-1} \, y^2 + a^2 \Delta_0^{1/2} \rho_0^{-4} \, y^1 y^2 , \\
x^2 & = r_0 \rho_0^{-1} \, y^3 + a^2 \Delta_0^{1/2} \rho_0^{-4} \, y^1 y^3 , \\
\begin{split}
x^3 & = r_0 + \Delta_0^{1/2} \rho_0^{-1} \, y^1 + \frac{1}{2}M\bigl(r_0^2-a^2\bigr) \rho_0^{-4} \, \bigl(y^1\bigr)^2 \\
& \quad - M r_0^2 \rho_0^{-4} \, \Bigl[\bigl(y^2\bigr)^2 + \bigl(y^3\bigr)^2\Bigr],
\end{split}
\end{align}
\end{subequations}
in which the metric takes the form 
\beq
\label{metric-local}
{\diff s}^2 = -(1+2gy^1)\bigl(\diff y^0\bigr)^2 + \bigl(\diff y^1\bigr)^2 + \bigl(\diff y^2\bigr)^2 + \bigl(\diff y^3\bigr)^2
\eeq
up to second-order corrections in the coordinates $y^i$.
Here $\rho_0=r_0^2+a^2$, $\Delta_0=r_0^2-2Mr_0+a^2$ and $g = M (r^2_0 - a^2)\rho^{-3}_0\Delta_0^{-1/2}$.

In the local coordinates $(y^0, y^i)$ the self-force for a static particle on the symmetry axis is
\beq
\label{force-local}
f^i = \frac{Me^2r_0}{(r_0^2 + a^2)^2} \delta^i_1 ;
\eeq
the absolute value of~\eqref{force-local} is
\beq
\label{module-local}
|f| = \frac{Me^2 r_0}{(r_0^2 + a^2)^2} .
\eeq
To check whether~\eqref{force-bl} and~\eqref{force-local} are in agreement, we transform our result to the same set of local coordinates~\eqref{local-coord} and then place the particle on the symmetry axis of the Kerr black hole.
However, we can immediately point out that~\eqref{module-local} is equivalent to our result~\eqref{module-generic} evaluated at $r=r_0$, $\theta=0$.

The transformation of~\eqref{force-bl} from Boyer-Lindquist into Cartesian-like coordinates~\eqref{cartesian-like-coord} yields the following non-zero components:
\begin{subequations}
\label{force-cartesian}
\begin{align}
\begin{split}
f_1 & =  \frac{Me^2 \sin\theta}{\Delta \rho^4} \biggl(1-\frac{2Mr}{\rho^2}\biggr)^{-1/2}  \\
& \quad \cdot \bigl[\Delta \bigl( a\sin\phi + r\cos\phi \bigr) - a^2 r \sin^2\theta \cos\phi \bigr] ,
\end{split} \\
\begin{split}
f_2 & =  \frac{Me^2 \sin\theta}{\Delta \rho^4} \biggl(1-\frac{2Mr}{\rho^2}\biggr)^{-1/2}  \\
& \quad \cdot \bigl[\Delta \bigl( a\cos\phi - r\sin\phi \bigr) + a^2 r \sin^2\theta \sin\phi \bigr] ,
\end{split} \\
f_3 & =  \frac{Me^2 \cos\theta}{\Delta \rho^4} \biggl(1-\frac{2Mr}{\rho^2}\biggr)^{-1/2} \bigl(\Delta - a^2 \sin^2\theta\bigr) .
\end{align}
\end{subequations}
It is worth noting that for $r=r_0$, $\theta=0$ only the component along the symmetry axis is non-zero
\beq
f_3 =  \frac{Me^2}{(r_0^2 + a^2)^2} \biggl(1-\frac{2Mr}{r_0^2 + a^2}\biggr)^{-1/2} .
\eeq
Finally, we move to the local coordinates $(y^0, y^i)$, in which we have
\begin{subequations}\label{local_force_y}
\begin{align}
\begin{split}
f_1 & =  \frac{Me^2r_0}{{\rho_0}^8} \biggl(1-\frac{2Mr_0}{{\rho_0}^2}\biggr)^{-1/2}  \\
& \quad \cdot \bigl[\sqrt{\Delta_0} {\rho_0}^3 + M \bigl(r^2 - a^2\bigr) y^1 \bigr] ,
\end{split} \\
f_2 & = -\frac{2M^2e^2{r_0}^3}{{\rho_0}^8} \biggl(1-\frac{2Mr_0}{{\rho_0}^2}\biggr)^{-1/2} y^2  , \\
f_3 & = -\frac{2M^2e^2{r_0}^3}{{\rho_0}^8} \biggl(1-\frac{2Mr_0}{{\rho_0}^2}\biggr)^{-1/2} y^3  .
\end{align}
\end{subequations}
From~\eqref{local-coord} we see that the position $r=r_0$, $\theta=0$ corresponds to
\beq \label{local1}
y^1 = y^2 = y^3 = 0 ,
\eeq
or
\beq \label{local2}
\begin{aligned}
y^1 &= -\frac{2\Delta_0^{1/2}{\rho_0}^3}{M(r^2_0-a^2)} ,\\
y^2 &= y^3 = 0.
\end{aligned}
\eeq
On the symmetry axis only the first component of the force is non-zero; indeed, using~\eqref{local1}, \eqref{local_force_y} simplifies to:
\beq
f_1 = \frac{Me^2r_0}{(r_0^2 + a^2)^2} .
\eeq
This expression is equivalent to~\eqref{force-local}, as can be easily checked by raising the index with the metric~\eqref{metric-local}.
We note that the position~\eqref{local2} corresponds to the symmetric position on the $z$-axis with respect to the black hole.

\bibliographystyle{unsrt}
\bibliography{Bibliography.bib}

\end{document}